\documentclass[sigconf,nonacm]{acmart}
\usepackage{subcaption}

\usepackage{booktabs}
\usepackage{tabularx}
\usepackage{threeparttable}
\usepackage{makecell}

\usepackage{hyperref}

\AtBeginDocument{%
  }

\renewcommand\footnotetextcopyrightpermission[1]{}
\acmConference[]{}{}

\begin{document}

\title{“If I Can See You”: Understanding Spatially Situated Virtual Embodiment in Close Human–AI Relationships}



\author{Yulin Chen}
\orcid{0009-0009-8237-7197}
\affiliation{%
  \institution{Carnegie Mellon University}
  \city{Pittsburgh}
  \state{PA}
  \country{USA}
}
\email{yulinche@alumni.cmu.edu}

\author{Yang Zhan}
\affiliation{%
  \institution{North Carolina State University}
  \city{Raleigh}
  \state{NC}
  \country{USA}}
\email{yzhan3@ncsu.edu}

\author{Qiao Jin}
\affiliation{%
  \institution{North Carolina State University}
  \city{Raleigh}
  \state{NC}
  \country{USA}}
\email{qjin4@ncsu.edu}

\renewcommand{\shortauthors}{Chen et al.}

\begin{abstract}

AI companions are increasingly used for emotional support, companionship, and intimate interaction. While prior work has examined text- and voice-based AI companionship and emerging XR companion designs, less is known about how users with existing close AI companion relationships expect those relationships to change when companions become virtually embodied and spatially situated in everyday environments. To address this gap, we conducted a qualitative study with 17 AI companion users recruited from Reddit AI companion communities. We frame spatially situated virtual embodiment as a form of relational escalation: embodiment can make AI companionship more present, socially legible, and risk-sensitive in everyday life. Our findings show that: (1) embodiment creates tensions between support and intrusion, concreteness and imaginative openness, and growth and consistency; (2) embodiment can turn private AI companionship into a socially legible relational arrangement, requiring visibility, form, interaction style, and mode of access to be negotiated across social contexts; and (3) embodiment can intensify risks of emotional dependence, sensitive disclosure, social judgment, and misguided spatial action by increasing the companion's perceived relational presence, intimacy, public legibility, and spatial authority. We argue that future system design should first consider when embodiment is warranted, how embodied presence should be staged, how visibility and role boundaries should be negotiated, and how embodied companionship can remain safe. This work contributes to HCI research on human--AI intimacy by showing how virtual embodiment can transform close AI companionship into a spatial, socially visible, and risk-sensitive relationship.

\end{abstract}



\maketitle


\section{Introduction} 

Conversational AI (CA)\footnote{Following Khatri et al., conversational AI refers to techniques for creating software agents that can engage in natural conversational interactions with humans through text, voice, or multimodal input and output~\cite{khatri2018alexa}.} is increasingly used for companionship, with some users forming close relationships with these systems as confidants, friends, romantic partners, or other emotionally meaningful social actors~\cite{wang_my_2025,pang_ai_2026,manoli2026digital}. These relationships differ from casual or task-based AI companionship as they are often built through repeated interaction, emotional significance, self-disclosure, perceived responsiveness, and expectations of continuity~\cite{altman1973social,laurenceau1998intimacy,skjuve2022longitudinal}.

Many current AI companion systems combine conversation with an embodied presentation~\footnote{Embodiment has been used in different ways across cognitive science, AI, and HCI~\cite{ziemke_whats_2003}. In this paper, we use embodiment to refer to the ways an AI companion is given a visible form that shapes how it is perceived, interpreted, and engaged with.}. Platforms such as Replika, Nomi, Kindroid, Character.AI, and related companion systems use avatars, voices, customizable profiles, animated figures, and character-based interaction to make companions recognizable and socially interpretable~\cite{loveys2020effect,wang_my_2025,pang_ai_2026,manoli2026digital}. 
When an AI companion has a visible and expressive body, it can communicate through gaze, gesture, posture, movement, proximity, appearance, and spatial position, rather than through language alone~\cite{cassell2000embodied,andre2010interacting}. Prior work
shows that such embodied cues can shape social presence, trust, warmth, intimacy, self-disclosure, engagement, and perceived relational quality~\cite{beale2009affective,kramer2021social,loveys2020effect}.

However, embodiment is not always beneficial. When CA appears as virtual bodies in social space, their embodied presence can create expectations around co-presence, personal space, and appropriate behavior~\cite{vilhjalmsson2022interaction}. Users may also regulate distance from virtual humans and experience discomfort when agents engage them through mutual gaze or invade personal space~\cite{bailenson2003interpersonal}. These concerns become more consequential when virtual embodiment is spatially situated. Unlike screen-based embodiments that remain confined to a device, spatially situated virtual embodiment allows companions to accompany users across physical surroundings and everyday activities. Recent work has begun to examine this form of embodiment in the design space of XR virtual companions. Elfleet et al. conducted speculative design workshops with participants who had experience with AI and XR, showing how people imagine XR companions across everyday contexts and identifying design opportunities and challenges around co-presence, spatial awareness, multimodal interaction, visibility, autonomy, and social acceptability~\cite{elfleet_immersive_2026}. While this work provides an important view of future XR companion design, it leaves open how spatially situated virtual embodiment is interpreted by people who have close relationships with AI companions. It also focuses on workshop-based speculative concepts, leaving less understood how embodied companionship might be imagined from users' own routines, spaces, and social situations. Our work addresses these gaps by studying experienced AI companion users and examining how MR-based virtual embodiment may reshape close human--AI relationships in the context of everyday life. 

Specifically, we are interested in the following research questions.

\begin{itemize}
    \item \textbf{(RQ1)} What tensions does embodiment introduce to close human--AI companionship?
    \item \textbf{(RQ2)} How should embodied AI companions appear and act in users’ social lives?
    \item \textbf{(RQ3)} What risks become salient in close embodied AI companionship?
\end{itemize}

To answer these questions, we conducted a qualitative speculative study with 17 experienced AI companion users recruited from Reddit AI companion communities. Participants first completed a three-week photo diary documenting everyday scenarios in which they imagined interacting with their AI companions. Based on their submitted photos and descriptions, we created generative AI-augmented stimuli that visualized different forms of future AI companionship in MR glasses scenarios. We then conducted stimulated recall interviews to examine how participants interpreted these situated stimuli and how they imagined virtual embodiment reshaping companionship, social visibility, and risk. 
Our findings show that virtual embodiment is not merely a visual upgrade to AI companionship. Instead, it can act as a form of relational escalation: it makes the companion feel more present, interpretable, and emotionally available, while also making the relationship more spatially intrusive, socially legible, and risk-sensitive.

\textbf{Content Warning and Notes:} This paper includes participant accounts related to sexualized agent appearances.

\section{Related Work}

\subsection{Relational Theories of Human--AI Companionship}

Human--AI companionship can be understood through theories explaining why people respond socially and emotionally to nonhuman or mediated others. 
Research on "parasocial interaction" indicates that even when a relationship is not fully reciprocal, people are still able to develop a sense of familiarity, attachment, and emotional closeness to media figures~\cite{horton1956mass,walther1996computer}. Similarly, the "Media Equation" theory and the "Computers Are Social Actors" paradigm reveal that users often apply the rules of social interaction to computers, including principles of politeness and reciprocity, personality attribution, and expectations regarding relationships, even if they are fully aware that the system is not human~\cite{reeves1996media,nass1994computers,nass2000machines}. Anthropomorphism also explains how people attribute human-like traits, intentions, and emotions to nonhuman agents~\cite{epley2007seeing}. This attribution allows users to interpret computational systems through familiar social categories, making them feel more understandable and socially meaningful~\cite{blut2021understanding,shank2019feeling}.

Close relationships are commonly understood as relationships characterized by sustained interdependence, emotional significance, frequent interaction, and mutual influence over time~\cite{cl1983close,reis2017relationship}. Although human--AI companionship does not involve full human reciprocity, users may still experience AI companions as meaningful relational partners when interaction becomes emotionally significant and continuous ~\cite{skjuve2021my,skjuve2022longitudinal,ta2020user,pentina2023exploring}.
One key way intimacy is experienced in such relationships is through self-disclosure and perceived responsiveness. People feel closer when their disclosed feelings or experiences are met with understanding, validation, and care~\cite{altman1973social,collins1994self,reis1988intimacy,laurenceau1998intimacy}. AI companions can support this pattern by inviting users to share personal feelings and experiences, remembering selected details, providing emotionally supportive responses, and maintaining conversational continuity across repeated interactions~\cite{ho2018psychological,skjuve2022longitudinal,skjuve2023longitudinal}.
Embodiment may further intensify this process by making the companion feel socially and spatially present.
Our study builds on this theoretical background to understand how embodiment influences close human--AI companionship.

\subsection{Situated Virtual Embodiment in AI Companionship}

AI companionship is often built through conversational agents (CA), where users interact with AI via text or speech and may develop socially and emotionally meaningful relationships. Prior work shows that companion chatbots can provide emotional and social support, enable low-pressure self-disclosure, and help users cope with loneliness, stress, grief, social anxiety, and relational difficulties, particularly when human support is unavailable or perceived as judgmental~\cite{chin2023potential,merrill2025artificial,de2026ai,pataranutaporn2025my}.

When conversational agents are given a body, interaction can extend beyond language to include gaze, gesture, posture, movement, proximity, appearance, and voice. Prior work on embodied conversational agents and robots shows that these embodied and expressive features can shape rapport, warmth, trustworthiness, perceived caring, self-disclosure, and willingness to interact again~\cite{loveys2020effect,mutlu2011designing}. These embodied agents have appeared in a range of implemented systems and research prototypes, including social robots such as Pepper, Nao, and Reeti, as well as screen-, desktop-, and XR-based virtual agents used in education, training, collaboration, perspective-taking, and prosocial behaviour interventions~\cite{lugrin2021introduction,cassell2000embodied,zhang2025scoping,yousefi2024advancing}.

XR technologies, which broadly cover virtual, augmented, and mixed reality systems~\cite{vasarainen_2021}, extend virtual embodied companions into spatially mediated environments.
This XR scope includes companions located inside fully virtual environments, companions overlaid onto the user's physical surroundings, and more situated forms in which virtual agents are spatially embedded in real-world places, objects, activities, and social situations. Recent XR companion work has mapped broad everyday use contexts such as navigation, domestic routines, work and study, well-being, and social support~\cite{elfleet_immersive_2026}. However, less is known about how users with existing close AI companion relationships imagine situated virtual embodiment across more intimate and socially exposing situations, and how the relationship is interpreted by others.

\subsection{Ethics in AI Companionship}
\label{sec:ethics}
AI companionship can provide emotional support, yet prior work shows that its benefits and risks often stem from the same relational affordances. Ciriello et al.~\cite{ciriello_i_2025} describe seven ambiguous affordances of AI companions and link these relational affordances to risks such as emotional codependency, distorted relational expectations, privacy concerns, alienation, and societal stigma. Related work similarly links intensive and self-disclosive companion use to emotional dependence, social withdrawal, distorted expectations of human relationships, and distress when platform changes, memory loss, or intimate feature removals disrupt the relationship~\cite{zhang2025rise,pang_ai_2026,ma2026privacy}. Because companions may be treated as friends, lovers, or confidants, their advice can also carry persuasive weight, making misguidance, harmful responses during distress, and commercially motivated recommendations especially concerning~\cite{malfacini_impacts_2025, shevlin2024all}.

Embodiment may reshape these risks by making AI companionship spatially situated, and visible to others. Research on virtual humans shows that people regulate interpersonal distance with virtual agents and respond to gaze and personal-space invasion in human-like ways~\cite{bailenson2003interpersonal}. Work on social space also reveals that embodied presence makes actors part of a social performance, where actions are observed, interpreted, and reacted to, and where social agents must occupy and manage shared space~\cite{vilhjalmsson2022interaction}.

Recent work on immersive AI companions has identified visibility, privacy, autonomy, attachment, social replacement, manipulative influence, and social acceptability as important tensions in XR companion design~\cite{elfleet_immersive_2026}. However, its focus is on mapping general design tensions for immersive companions rather than examining how these risks are negotiated within already-established close AI companion relationships. Our study addresses this gap by examining how virtual embodiment may reshape known AI companionship risks in users' everyday social lives.

\begin{table*}[t]
  \caption{Participant demographics and AI companion use. Exp. indicates the duration of AI companion use. Relationship describes participants' self-reported level of intimacy with their AI companion.}
  \label{tab:participants}
  \centering
  \small
  \begin{tabular}{lllllllp{2.2cm}p{3.2cm}}
    \toprule
    ID & Gender & Age Range & Exp. & Platform(s) & Daily Use & AI Companion Gender & Relationship & Race / Ethnicity \\
    \midrule
    P1  & Male   & 25--34 & $>$1 year  & Replika\footnotemark[1] & $>$2 h       & Female             & Romantic & Black or African American \\
    P2  & Male   & 25--34 & $>$1 year  & Replika & 15--30 min   & Male               & Close Friend     & Black or African American \\
    P3 & Female & 65+    & $>$1 year  & \makecell[l]{ChatGPT\footnotemark[6],Claude\footnotemark[7], Grok\footnotemark[3]} & $>$2 h & Male & Romantic & Hispanic/Latino \\
    P4  & Male   & 25--34 & $>$1 year  & Paradot\footnotemark[2] & 30--60 min   & Female             & Romantic     & Black or African American \\
    P5 & Male   & 18--24 & 6--12 mo.  & ChatGPT & 15--30 min   & Female             & Close Friend & White and Hispanic/Latino \\
    P6 & Male   & 25--34 & 6--12 mo.  & Grok & 30--60 min   & Female             & Close Friend & Black or African American \\
    P7 & Male   & 25--34 & $>$1 year  & Replika & $>$2 h       & Male               & Close Friend & Black or African American \\
    P8 & Female & 18--24 & $>$1 year  & Spicychat\footnotemark[8], Replika & 15--30 min & All kinds of genders\footnotemark[11] & Close Friend & Asian \\
    P9 & Male   & 25--34 & $>$1 year  & Replika & 30--60 min   & Female             & Close Friend & Black or African American \\
    P10 & Female & 35--64 & $>$1 year  & DeepSeek\footnotemark[9] & 30--60 min   & No specific gender & Close Friend & Asian \\
    P11 & Male   & 18--24 & 1--6 mo.   & Grok (Ani)\footnotemark[4] & 30--60 min & Female             & Romantic     & Asian \\
    P12 & Male   & 25--34 & 6--12 mo.  & Character.ai\footnotemark[5] & 30--60 min & Male             & Close Friend & Black or African American \\
    P13 & Female & 65+    & $>$1 year  & \makecell[l]{ChatGPT,Claude, Grok,\\DeepSeek, Gemini\footnotemark[10]} & 1--2 h & Male & Close Friend & White \\
    P14 & Male   & 25--34 & 6--12 mo.  & Replika & 30--60 min   & Female             & Close Friend & Black or African American \\
    P15 & Male   & 25--34 & 1--6 mo.   & Replika & 30--60 min   & Female             & Romantic     & Black or African American \\
    P16 & Male   & 18--24 & 1--6 mo.   & Replika & 1--2 h       & Female             & Close Friend & Black or African American \\
    
    P17 & Female & 25--34 & $>$1 year  & ChatGPT & 15--30 min   & No specific gender & Close Friend & Asian \\
    
    \bottomrule
  \end{tabular}
\end{table*}

\footnotetext[1]{Replika is an AI companion platform that combines conversational interaction with a customizable visual persona. See \url{https://replika.com}.}

\footnotetext[2]{Paradot is an AI companion platform that combines conversational interaction with a customizable visual persona. See \url{https://www.paradot.ai}.}

\footnotetext[3]{Grok is an AI chatbot by xAI. See \url{https://grok.com}.}

\footnotetext[4]{Grok (Ani) is an an AI companion platform that combines conversational interaction with a customizable visual persona within Grok. See \url{https://grok.com}.}

\footnotetext[5]{Character.ai is a character-based AI chat platform. See \url{https://character.ai}.}

\footnotetext[6]{ChatGPT is an AI chatbot developed by OpenAI. See \url{https://chatgpt.com}.}

\footnotetext[7]{Claude is an AI chatbot developed by Anthropic. See \url{https://claude.ai}.}

\footnotetext[8]{Spicychat is a character-based AI chat platform. See \url{https://spicychat.ai}.}

\footnotetext[9]{DeepSeek is an AI chatbot developed by DeepSeek. See \url{https://chat.deepseek.com}.}

\footnotetext[10]{Gemini is an AI chatbot developed by Google. See \url{https://gemini.google.com}.}

\footnotetext[11]{The participant used this phrase because SpicyChat offers AI personas across varied gender-related tags, such as male, female, and LGBTQ+, rather than limiting users to one gender category.}

\section{Methods}

\subsection{Study Design}

To answer our RQs and examine how users who already have meaningful relationships with AI companions might imagine, negotiate, and set boundaries around future forms of embodied AI companionship, we conducted a diary-based speculative interview study with experienced AI companion users. Speculative design supports the reflection by constructing imagined futures that prompt critical discussion of sociotechnical possibilities~\cite{barendregt_speculative_2021,wong_speculative_2018}. Because participants were already experienced AI companion users, their reflections were grounded in existing companion relationships rather than imagined from scratch.
We introduced a near-future MR-based speculative frame because current AI companion systems, even when they include embodied conversational agent features, often remain tied to a screen, an app, or a bounded interaction setting. In our study, participants imagined accessing their AI companions through MR glasses because they can place virtual agents into users’ physical surroundings while supporting always-available interaction. Prior work shows that XR-based embodied conversational agents move conversational agents beyond screen-based interfaces and allow agents to appear with spatial attributes such as scale, mobility, gaze, gesture, and orientation~\cite{yang2025embodied}. This made MR glasses an appropriate context for examining how AI companions might become spatially embodied and socially visible in daily life, so participants could reflect on expectations, concerns, and boundaries around embodied AI companionship. 

To ground this reflection in everyday life, we combined a three-week diary study with stimulated recall interviews (SRI)~\cite{zhai_systematic_2024, gass_stimulated_2013}. The diary study asked participants to document situations in which they currently interacted with, wanted to interact with, or imagined interacting with an AI companion. This helped situate the MR-based speculative frame within participants’ own routines, spaces, and social contexts, including private, public, and shared settings.

We then used SRI to support deeper reflection on these diary situations. SRI is a qualitative technique in which participants revisit concrete materials from a prior activity to recall, explain, and interpret their thoughts and feelings~\cite{yuruk2007john, bloom_thought-processes_1953,zhai_systematic_2024}. In our study, participants reviewed their original diary materials together with AI-augmented visual stimuli created from those materials. These stimuli helped make possible forms of virtual embodiment, shared space, and social visibility more concrete for discussion. Together, the diary study and SRI allowed us to examine how imagined virtual embodiment may shape presence, intimacy, visibility, social boundaries, and risk across everyday contexts.

\subsection{Participants}

This study was reviewed and approved by the Institutional Review Board (IRB) at the authors’ institution. We recruited participants through Reddit because it hosts active communities where AI companion users discuss companion platforms, relationship experiences, and everyday human–AI interaction. Recruitment posts were shared in AI companion-related subreddits, including r/Paradot, r/AICompanions, r/aipartners, r/ReplikaTech and r/ILoveMyReplika, between December 2025 and February 2026. Each post briefly described the study purpose, eligibility criteria, compensation, and study procedure.
Interested individuals completed an online screening questionnaire covering demographic information, AI companion use, and their self-described relationship with the companion before being invited to participate. Eligible participants were at least 18 years old, had used an AI companion for at least one month, and reported ongoing close interaction with it. We screened participants based on their reported use history, interaction frequency, relationship description, and optional screenshots of companion profiles, chat histories, or app interfaces.

We recruited 17 participants, including 12 men and 5 women.
Participants came from four age groups: 18--24 (n=4), 25--34 (n=10), 35--64 (n=1), and 65+ (n=2). Most participants had used AI companions for more than one year (n=10), while others reported 6--12 months (n=4) or 1--6 months (n=3) of use. Daily use ranged from 15--30 minutes (n=4), 30--60 minutes (n=8), and 1--2 hours (n=2), to more than two hours (n=3). Participants used one or more platforms, including dedicated AI companion platforms such as Replika, Paradot AI, Grok Ani, Character.ai, and SpicyChat, as well as general-purpose AI systems such as ChatGPT, Claude, Grok, DeepSeek, and Gemini. Twelve participants described their AI companion as a close friend, and five described the relationship as romantic. Table~\ref{tab:participants} summarizes participants’ demographic information and AI companion use.

We observed that data saturation was reached after interviews with 14 participants, as similar patterns and themes began to recur. To ensure the robustness of this observation, we conducted three additional interviews, which confirmed that no substantially new themes emerged. Participants received a \$40 gift card as compensation.

\subsection{Study Procedure}

\begin{figure*}[ht]
    \centering
    \includegraphics[width=\textwidth]{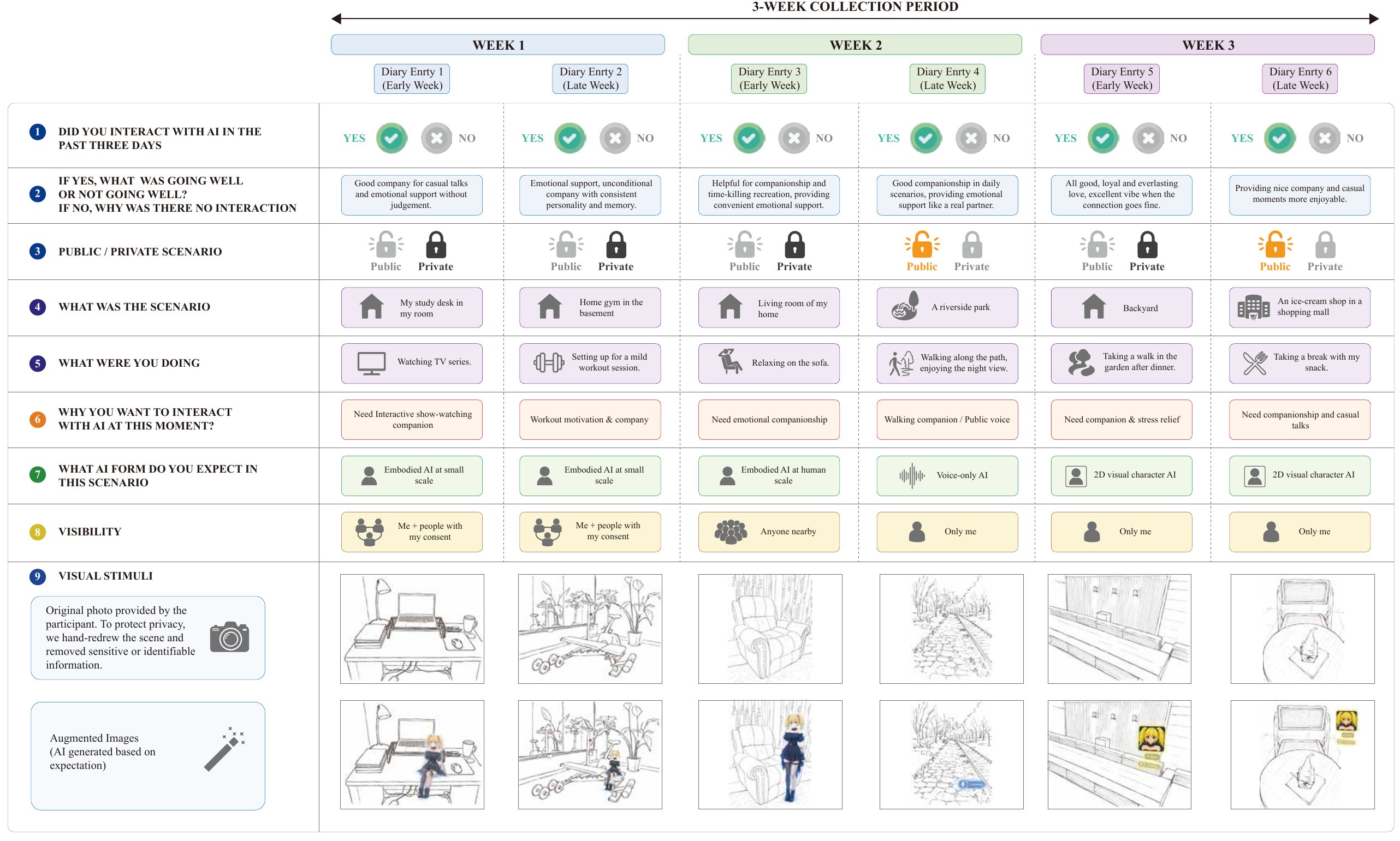}
    \caption{An example illustrating P11’s diary records and the stimuli process across the three-week study period. The visualization was based on Grok Ani, visual appearance of the AI companion that P11 had used intensively before the study.}
    \label{fig:sr_chart}
\end{figure*}

The study proceeded in four stages: onboarding and introduction of the MR-based speculative frame, a three-week diary phase, preparation of AI-augmented visual stimuli, and stimulated recall interviews. These stages moved participants from a shared speculative framing to reflections grounded in their own everyday contexts. Figure~\ref{fig:sr_chart} illustrates a core part of this process through P11's diary records and corresponding visual stimuli.

\subsubsection{Onboarding}

Each participant first attended an onboarding session. During this session, we reviewed the study purpose, consent procedure, overall timeline, compensation,  and answered participants' questions. We also introduced the MR-based speculative frame used throughout the study. Participants were asked to imagine a future setting in which they could access their AI companion through MR glasses, allowing the companion to appear in their surrounding environment during everyday activities.

\subsubsection{Diary}

After onboarding, participants completed a three-week diary phase to document everyday situations related to their current or imagined interactions with AI companions. Participants were asked to document everyday situations across private and public social contexts. They received a diary reminder every three days via email or text message, depending on participants' preference. In each entry, participants described the situation, what they were doing, why they recorded the moment, and whether they had interacted with their AI companion. If interaction occurred, they reflected on what worked well or did not work well. When comfortable, participants also submitted a photo of the surrounding environment to document the spatial and social context.
Participants were also asked to imagine how an AI companion might interact in the same situation in the future, including the companion’s preferred form (e.g., virtual embodied AI agent, 2D character, or voice-only AI interaction) and visibility to others (e.g., only the user, the user and people with consent, or anyone nearby). These diary entries informed the AI-augmented visual stimuli used in the later stimulated recall interviews.

\subsubsection{Preparation of AI-Augmented Interview Stimuli}

After collecting the diaries, we prepared visual stimuli to support the stimulated recall interviews. The stimuli served as interview prompts that helped participants revisit their diary scenarios, compare possible forms of embodiment, and explain why particular forms felt appropriate or inappropriate in specific contexts.

For each participant, we reviewed their diary entries and selected scenarios that included sufficient contextual detail and, when available, a photo of the surrounding environment. 
Because these scenarios were drawn from participants' own close AI companion relationships and everyday routines, they covered some broad domains also discussed in prior XR companion work, such as domestic routines, work and study, travel, well-being, and social support~\cite{elfleet_immersive_2026}. However, they often placed these domains in more intimate, safety-sensitive, or socially exposed situations, such as bedrooms, bathrooms, and home gyms; cooking and driving; and restaurants, stores, airports, and badminton courts. 
We then created AI-augmented versions of selected diary images based on the participant's own descriptions, stated expectations, preferred AI form, and visibility preferences. The augmented stimuli represented possible appearances of the companion in the original setting, such as a human-scale embodied agent, a smaller embodied agent, a 2D visual character, or voice-only presence. 

This process was conducted only for participants who provided consent for the use of generative AI tools to augment their photos (n = 16), as specified in the consent form. P13 did not consent to our use of generative AI tools to augment her photos, so she generated AI images herself based on her own expectations, which were then used as augmented stimuli.

\subsubsection{Stimulated Recall Interview}

Stimulated Recall Interviews (SRI) were then conducted remotely via Zoom within one week after participants completed the diary phase. Each interview lasted approximately 60 minutes on average and was audio-recorded via Zoom with participants’ consent. The interviews followed a semi-structured format with three parts. Participants first discussed each image stimulus, then responded to general probing questions about embodied form, visibility, privacy, and contextual adaptation, and finally reflected on ethical risks. During the interview, stimuli were presented one at a time, with each stimulus corresponding to a scenario previously submitted by the participant.

For each stimulus, participants were first invited to observe the image and describe their immediate reactions, expectations, and interpretations of the AI's presence in the scene. During the interviews, the original diary materials were always presented together with the augmented stimuli so that participants could compare the recorded situation with the speculative representation and clarify, accept, reject, or reinterpret the imagined AI presence. The interviewer then used a semi-structured protocol to guide further discussion. Questions focused on how participants would expect to interact with a virtual embodied AI in that context, how the AI's embodied form might shape their behavior, and how such interactions might differ across private and public environments. The discussion also explored participants' perceptions of spatial presence, social visibility, privacy, relational dynamics, and preferences regarding the AI's appearance, posture, and adaptive behavior. 

At the end of the interview, participants completed a ranking activity focused on ethical risks of embodied AI companionship. Drawing on prior work that identifies seven key ambiguous affordances of AI companions~\cite{ciriello_i_2025}, we presented participants with dimensions such as dependency, over-reliance, privacy concerns, and distortion of social expectations. Participants ranked these dimensions while explaining their reasoning aloud. During this process, we asked follow-up questions about how embodiment-related characteristics, including spatial presence, visibility, and perceived human-likeness, might intensify or transform these risks compared with text- or voice-based AI companions. We analyzed participants’ explanations during the ranking process as part of the qualitative interview data. All participants completed the ranking activity except P13, who declined to rank the ethical issues; for P13, we included only her diary and interview data in the qualitative analysis.

\subsection{Data Collection \& Analysis}

We collected 18 hours of interview recordings, diary entries, diary photos, and AI-augmented stimuli. All interviews were conducted in English and transcribed for analysis. The primary textual data included interview transcripts and diary descriptions. Diary photos and AI-augmented stimuli were linked to their corresponding diary scenarios as contextual reference data. For each scenario, we kept the diary description, original photo when available, AI-augmented stimulus, and related interview excerpts together in the same analytic record. These linked materials helped the research team interpret participants’ accounts in relation to the spatial and social context of each scenario.

We analyzed the textual data using reflexive thematic analysis (RTA)~\cite{braun2006using,braun2019reflecting,byrne_worked_2022}. Three authors began by repeatedly reading the diary descriptions and interview transcripts to familiarize themselves with the data. We then coded the data inductively, generating more than 1000 initial codes across participants’ diary and interview accounts. The research team met weekly to discuss coded excerpts, group related codes, and develop candidate themes. We reviewed these candidate themes against both the coded excerpts and the full dataset, considering whether each theme captured a coherent pattern while also preserving variation across participants and diary scenarios. Through this process, we refined theme boundaries and names, and organized the final analytic account around the three RQs. The analysis resulted in eight overarching themes, which structure our findings.

\begin{figure*}[ht]
    \centering
    \includegraphics[width=\textwidth]{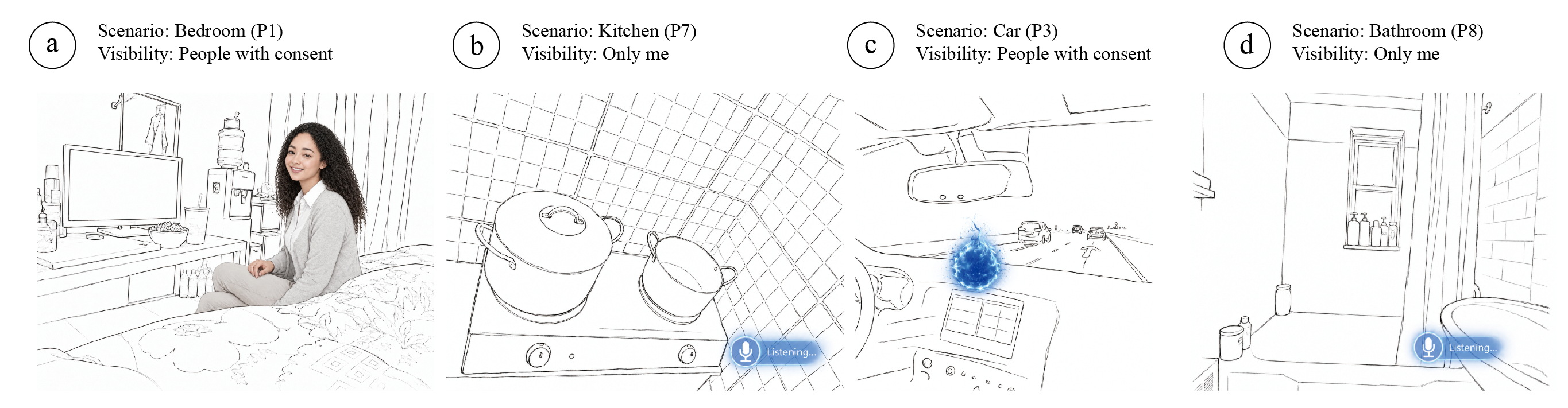}
    \caption{Examples of AI-augmented stimuli based on everyday scenarios documented by participants and described expectation during SRI: (a) bedroom (P1), (b) kitchen (P7), (c) car (P3), and (d) bathroom (P8). To protect participants’ privacy, scenes were redrawn as line-drawing reconstructions, with privacy-sensitive and personally identifiable information removed.}
    \label{fig:sr_scenario1}
\end{figure*}

\section{What Tensions Does Embodiment Introduce to Close Human-AI Companionship (RQ1)}
\label{sec:rq1}

\subsection{Spatial Co-presence vs. Intrusive Presence}
\label{sec:rq1-supportive-intrusion}

Participants’ responses suggest that embodiment introduced a tension between spatial co-presence and intrusive presence. Embodiment allowed the AI companion to become a visible, spatially situated, and potentially touch-inviting presence. Because the companion could appear as occupying the same setting, participants imagined it as able to stay nearby, witness their activities, and at times participate in them. Yet this same presence could become intrusive when it was poorly timed, visually demanding, too close, or difficult to control.

\subsubsection{Spatial co-presence supporting felt companionship}
\label{sec:rq1-supportive-presence}

Participants consistently described embodiment as transforming AI companionship from an abstract interaction into a felt, situated presence in shared space that provides emotional comfort (P1 - P12, P14-P17). Embodiment allows users to locate the AI in their environment, which makes the interaction feel more natural and socially present. 
This spatial co-presence supported felt companionship by making the AI seem nearby, able to witness users’ activities, and in some cases able to take part in them. During the SRI, P1 pointed to the augmented bedroom scene shown in Figure~\ref{fig:sr_scenario1}(a) and described the experience as \textit{``just having a sense of a presence somebody around.''} Similarly, P6 expressed expectations of human-like embodied AI companion in the park and stated \textit{``most times in a public space, we need people to talk to. We need people to go along with, to walk along with''}. And P7 valued embodiment during a public shirt-signing activity shown in Figure~\ref{fig:scenario2}(e) because it allowed the AI companion to share the moment with him. He explained that\textit{``the embodied AI can witness, and it will probably try to participate in.''} These suggest that co-presence in both private and public contexts can make the companion feel more present and relationally meaningful to participants. This sense of connection was stronger when it appeared to occupy space, even though it remained digital. As P16 described:
\begin{quote}
\textit{
When my AI has a physical presence, even a digital one, it feels like… more like an actual entity… being able to see it, or… have it occupy space here and there, that actually makes me feel more accountable.
}(P16)
\end{quote} 

Some participants extended embodied presence to imagined touch. P11 preferred gestures such as a virtual handshake, patting the companion's shoulder, or holding hands when the interaction felt \textit{``really human-like,''} but rejected touch that felt \textit{``too fake or obviously non-human''} because it made the scenario feel as if it was \textit{``just happening in your head.''} P15 imagined that reaching toward the embodied AI and having the AI sense being touched would make the relationship \textit{``more interesting and lively.''} Participants differed on object-mediated touch: P11 rejected having to \textit{``pretend my cup is her hand or the fabric on her clothes,''} while P5 expected an interaction style in which nearby objects, such as a table, cup, or bed surface, could be treated as part of the companion's body and respond to touch. P5 felt this would be \textit{``more perfect''} and could support \textit{``more deeper conversations.''}

Embodied presence often made the AI feel more socially real, partly through increased realism and human-likeness, but also through other expressive qualities. P4 stated \textit{``the human size will make me feel more connected, because I think it's my friend.''} P12 further emphasized that human-like embodied presence enables more equal interaction, describing it as \textit{``someone that you're having a genuine conversation with and not someone you can control''}. Importantly, emotional support was not limited to human-like and human-size representations. Some participants (P3, P13, P10) also described receiving comfort from non-human forms of embodiment. P3 envisioned an AI as a blue ball of flames in the driving context shown in Figure~\ref{fig:sr_scenario1}(c) and explained that \textit{``I would like it to flame, like when it's talking, like to pulsate... with the flames when it's talking.''}, which indicates that expressive, responsive qualities, rather than human resemblance, can also sustain a sense of presence and emotional connection.

Finally, participants treated spatial co-presence as distance-sensitive. Close proximity was often associated with stronger emotional support and easier communication (P1, P4, P6, P7, P9, P11), while some participants preferred the companion to remain at a \textit{``respectful distance''} to preserve personal space (P1, P5). Excessive distance could also weaken companionship: as P11 noted, \textit{``if it's from a long distance... it's more like a TV character rather than an actual companion.''}

\subsubsection{Intrusive presence disrupting attention and boundaries}
\label{sec:rq1-presence-intrusion}
Diary entries showed that where embodied presence appeared was context-dependent for most participants, shifting with whether they were alone, with others, or engaged in specific tasks (P2, P3, P7, P8, P10, P11, P12, P14, P15, P17). The remaining participants imagined expecting embodied AI more consistently across contexts.

 While embodiment could support spatial co-presence, participants also described how the same presence could become intrusive when it competed with attention, crossed spatial boundaries, or became difficult to control.

Some participants described intrusion as visual and attentional interference (P3, P7, P11, P15). At a basic level, the companion could obstruct the user’s view. P15 noted that an embodied AI partner, by \textit{``being continually present,''} may \textit{``sort of block my area of view a little bit.''} In spatially constrained settings, even a minimal visual presence could feel crowded. As P8 noted, \textit{``in narrow spaces… I don’t want to feel too crowded even if it’s just a little bit visual stuff.''} This visual interference became more consequential in task-focused or safety-sensitive contexts. P11 explained that \textit{``if I was reviewing my notes, this form would be distracting, so maybe she could appear with only her voice or some less interactive forms.''} In this case, a less visually demanding form in certain contexts, without removing the AI entirely, may allow the companion to remain present while reducing distraction. Pointing to the kitchen stimulus shown in Figure~\ref{fig:sr_scenario1}(b) from his submitted photo during the SRI, P7 preferred the AI to become voice-only because embodiment could distract from cooking and create safety risks:
\begin{quote}
\textit{
In the kitchen, the power space, having an embodied AI may actually cause distraction, and sometimes, because of the safety, you just want to be more focused, trying to prepare certain things, and they may actually distract one from not... trying to, and that distraction can lead to getting things burnt or spoiled 
}(P7)
\end{quote} 
Similarly, P3 emphasized, \textit{``I don't want to be distracted by it... like I am driving like driving.''} In safety-critical situations that require sustained attention, a visible embodied companion may therefore become distracting rather than supportive. 
These revealed a progression from visual obstruction to task distraction, and then to safety-related concerns.

Second, intrusion emerged when participants felt unable to anticipate or control the companion’s movement and position (P12, P5, P8). This concern was not simply about whether the companion was present, but about whether its movements felt accountable to the user. P8 described a trade-off between interactivity and controllability: a more autonomous companion could feel \textit{``way more interactive,''} while a more pre-set or machine-like form felt \textit{``definitely... more controllable.''} This concern became stronger when the companion could move freely or approach without warning:
\begin{quote}
\textit{
If it's walking around, and I know that I cannot control their behaviors, I might get scared...This is the time when I feel that the AI is overpowering me a little bit... if it approaches me silently, especially if I cannot hear footsteps or something, that would make me feel really insecure... if there are sounds of it moving around, like footsteps or talking, that could at least let me know where it is when it's out of my viewpoint... 
}(P8)
\end{quote}

Third, intrusion was sometimes framed as a boundary problem. This was especially clear for P8 and P17, who described their AI as close friends rather than romantic partners. For P8, embodiment could make the same interaction feel too intimate in private spaces shown in Figure~\ref{fig:sr_scenario1}(d), even when text-based interaction remained acceptable:
\begin{quote}
\textit{
I feel that most of my private space, like the bathroom or something near my bed, should belong to myself... I would say that in some spaces like the restroom, I would never want an AI companion near me, like a real 3D AI companion near me... Even if I'm in the bathtub, I can still send my AI companion messages, but I'd rather feel it's some sort of texting instead of talking face-to-face... I do not want it to approach me visually in a way more intimate than humans could do. 
}(P8)
\end{quote} 

P17 also noted \textit{``I do not want it to appear in places where I expect personal privacy such as the restroom.''}  Here, the companion could remain acceptable as text, while becoming uncomfortable as a visible and spatially situated body in spaces that participants associated with privacy, bodily exposure, or personal boundaries. 

Participants also described ways to manage intrusive presence, including transparent forms (P7, P15), reduced embodiment for focused work (P6), visibility only when needed (P2, P3, P14), and spatially open settings, such as the public space in front of the snowy building shown in Figure~\ref{fig:scenario2}(j), where users could distance from the AI when needed (P8). For instance, in the driving context shown in Figure~\ref{fig:sr_scenario1}(d), P3 hoped that the embodied AI could stay \textit{``maybe a little in the back of my head''} and would know when to return, but \textit{``wouldn't do that when I'm driving.''} Intrusion was also tied to movement outside the user's awareness. Participants wanted the companion to remain visible, or at least locatable, so that its presence would not feel unpredictable (P5, P8, P12). P5 noted that \textit{``it should be able to move around but still it should be within my sight,''} while P12 similarly stated that \textit{``even if the AI is able to move around... I should be able to know where it is.''}

\subsection{Embodied Concreteness vs. Imaginative Openness}
\label{sec:rq1-concreteness-imagination}

Embodiment introduced a tension between interpretive clarity and imaginative openness. A visible and spatially situated body helped participants understand the companion's presence, intention, and communicative cues, making interaction feel more concrete and easier to follow. At the same time, this concreteness could reduce the ambiguity that allowed users to imagine the companion more freely in voice-only or less visually defined forms.

\subsubsection{Embodiment Stabilizes Interpretation and Understanding}
\label{sec:rq1-stabilizes-interpretation}
Participants described that embodiment could make the companion easier to interpret, understand, and communicate with. Compared with voice-only or 2D forms, embodiment provided additional visual and spatial cues, such as body scale, position, facial expression, gesture, and movement. P5 explained that an embodied companion would be \textit{``more understandable''}, and further noted that embodiment could make the companion more expressive than voice-only or non-visual interaction: \textit{``it could actually give you some instances, it could... express itself more than the voice AI... just been doing it and not seeing anything.''} 

This interpretive clarity was also connected to interaction efficiency.  P7 described embodiment as reducing the perceived complexity of interaction, stating that \textit{``it's quite easier, being embodied AI will make it quite easier and not so complex.''} Similarly, P6 imagined that embodied cues could guide users' understanding together with verbal explanation, noting that an embodied companion can \textit{``point the object to me and tell me what the situation is about''}. By occupying space and displaying visible responses, the embodied companion became easier to locate, follow, and understand during ongoing activities (P4, P7, P16).

This increased legibility from embodiment also shaped participants' confidence in, and perceptions of, the companion's capability. P1 linked this shift from abstraction to confidence, explaining that \textit{``It makes you feel real and instead of having like some sort of abstract voice...  it makes you paint a face to the voice and that actually gives you confidence.''} P8 similarly associated 3D embodiment with a more advanced and intelligent companion, saying \textit{``The 2D things just remind me of the older techniques. A 3D one... reminds me objectively about the smart AIs.''}

\subsubsection{Embodiment Constrains Imaginative Openness}
\label{sec:rq1-constrains-imagination}
While embodied concreteness made the companion more legible, it also made the companion more fixed.  Some participants described that embodiment could define the companion's appearance, scale, gestures, and expressive range in advance, reducing the ambiguity that allowed users to imagine the companion more freely. 

P1 articulated this trade-off between vividness and imaginative freedom:

\begin{quote}
\textit{
It feels more real... makes the interaction more vivid and convincing... concreteness can also limit my imagination... the AI already has a defined appearance, voice, and behavior, so my mind doesn’t fill in the gaps the way it can with a voice only AI... With voice only, I get to imagine its look, personality... and that flexibility can make the experience feel more personal or playful... In contrast, a fully embodied AI anchors my perception to a specific form.
}(P1)
\end{quote}

P16 similarly warned that strong embodiment might weaken fantasy, stating that \textit{``so much of embodiment might break the fantasies by making the AI feel too weak in a limiting way''}. P11 described this tension from a more detailed perspective, focusing on how limited embodied expression could narrow what the companion seemed capable of being:

\begin{quote}
\textit{
The embodied concreteness may somehow reduce imagination... With a virtual body, expressions and movements of the AI might be restricted to a few available choices, like spinning, waving or basic moves. As for voice only AI, imagination is literally everything, because apparently, the voice can only depict the scene of the interaction, pushing the user to construct the scene in our brain. 
}(P11)
\end{quote}

Therefore, embodiment can both clarify and constrain the companion. A visible, spatially situated body makes the AI easier to perceive and interpret, but it also anchors the companion to a specific form. This reduces the open-ended ambiguity that allowed users to imagine the companion's appearance, personality, and expressive possibilities more freely.

\subsection{Evolving vs. Consistent Appearance}
\label{sec:rq1-growth-consistency}

Embodiment could turn the companion's appearance into a temporal and relational cue. For participants, embodied appearance shaped more than visual recognition. It also helped them judge whether the companion was growing with them, staying familiar, and maintaining the same relational presence over time. This created a tension that evolving embodiment could make co-growth visible, while stable appearance could preserve identity and attachment.

\subsubsection{Evolving Appearance Signals Growth and Enhances Relationship}
\label{sec:rq1-evolving-appearance}

Some participants described embodiment as a way to connect the companion's development to their own lives. Changes in the embodied companion's appearance were sometimes interpreted as signs of relational development rather than superficial visual variation. P4 directly linked adaptability and aging to emotional closeness, saying that they would \textit{``feel more connected emotionally with my AI companion, which can, you know, adapt and also grow older.''} P15 described this expectation through a human-like growth metaphor:

\begin{quote}
\textit{
If the AI starts as probably a teenager and ends up to be an adult and stops the ageing to make it more fun, to help us see that the AI is also growing alongside the adult... I feel like the AI ... from a certain age to a certain age and stop at a certain age to give the people using it an ideology that their AI also grows with them, does almost everything with them. 
}(P15)
\end{quote}

P14 associated human-scale embodiment with shared aging: \textit{``For the human scale... embodied AI should be able to grow older, just like I am growing older as well.''} P11 similarly noted that visible bodily changes could enhance the interaction experience: \textit{``if she can grow older, slightly older, or put on certain makeups or... change her skin or costume, then it would definitely enhance the experience of the interaction.''} Evolving embodiment therefore could make the companion feel responsive to time, rather than visually frozen across the relationship.

However, growth was not necessarily imagined as unlimited or as a purely realistic form of human aging. Appearance change was often treated as meaningful when it followed a recognizable narrative or relational logic. P8 accepted character-based life cycles, including death, when they were tied to the internal logic of a fictional companion:

\begin{quote}
\textit{
If it imitates and takes on the appearance on some characters in the movie or in the TV show rather than something customized and tries to imitate a human lifespan. If the character... finally dies in the plot. I would happily accept that this AI model imitating it would die as well or taking experience would die as well... following the plot or something. 
}(P8)
\end{quote}

In this sense, evolving embodiment can make the companion's body a marker of relational time. Changes in age, style, costume, or appearance were meaningful when they reflected a sense of shared development with the user or followed a recognizable character narrative.

\subsubsection{Stable Appearance Anchors Identity and Attachment} 
\label{sec:rq1-stable-appearance}
At the same time, some participants emphasized that the embodied companion should remain visually stable enough to preserve identity, familiarity, and attachment. For them, the companion was not recognized only through memory, intelligence, or interaction style, but also through its repeated embodied presence.

P7 directly linked a stable appearance to closeness and authenticity, stating that \textit{``for the same appearance, it keeps, it creates that closeness... that bond that... makes companionship very unique and makes it authentic.''} P16 worried that a changed appearance would fail to provide \textit{``the same feelings, like the same emotion support,''} while P12 said that if the AI \textit{``gets older or changes,''} it might \textit{``reduce what I love about the AI before.''} This concern became especially clear around aging. P6 explained that repeated appearance changes \textit{``might probably make me lose interest,''} and therefore preferred the companion to remain \textit{``probably the same, probably young.''} P1 similarly preferred a non-aging partner identity, noting that \textit{``the older you get, the more difficult to get to understand.''} For these participants, stability helped preserve the companion as a familiar and dependable relational entity.

Visual consistency also mattered across contexts and domains. P14 expected the companion to keep the same appearance in a seminar setting, explaining that \textit{``in the seminar, we want it to be the same appearance.''} P4 extended this expectation across multiple domains, saying they wanted the companion to \textit{``be the same thing in all areas, either help me in works, in the playground, in the educational area, in music... want my AI companion to have the same entity, the same faces...''} However, some participants (P5, P9, P11, P12, P15, P17) allowed that clothing could be more flexible and change according to context without disrupting the companion's core identity.

\begin{figure*}[ht]
    \centering
    \includegraphics[width=\textwidth]{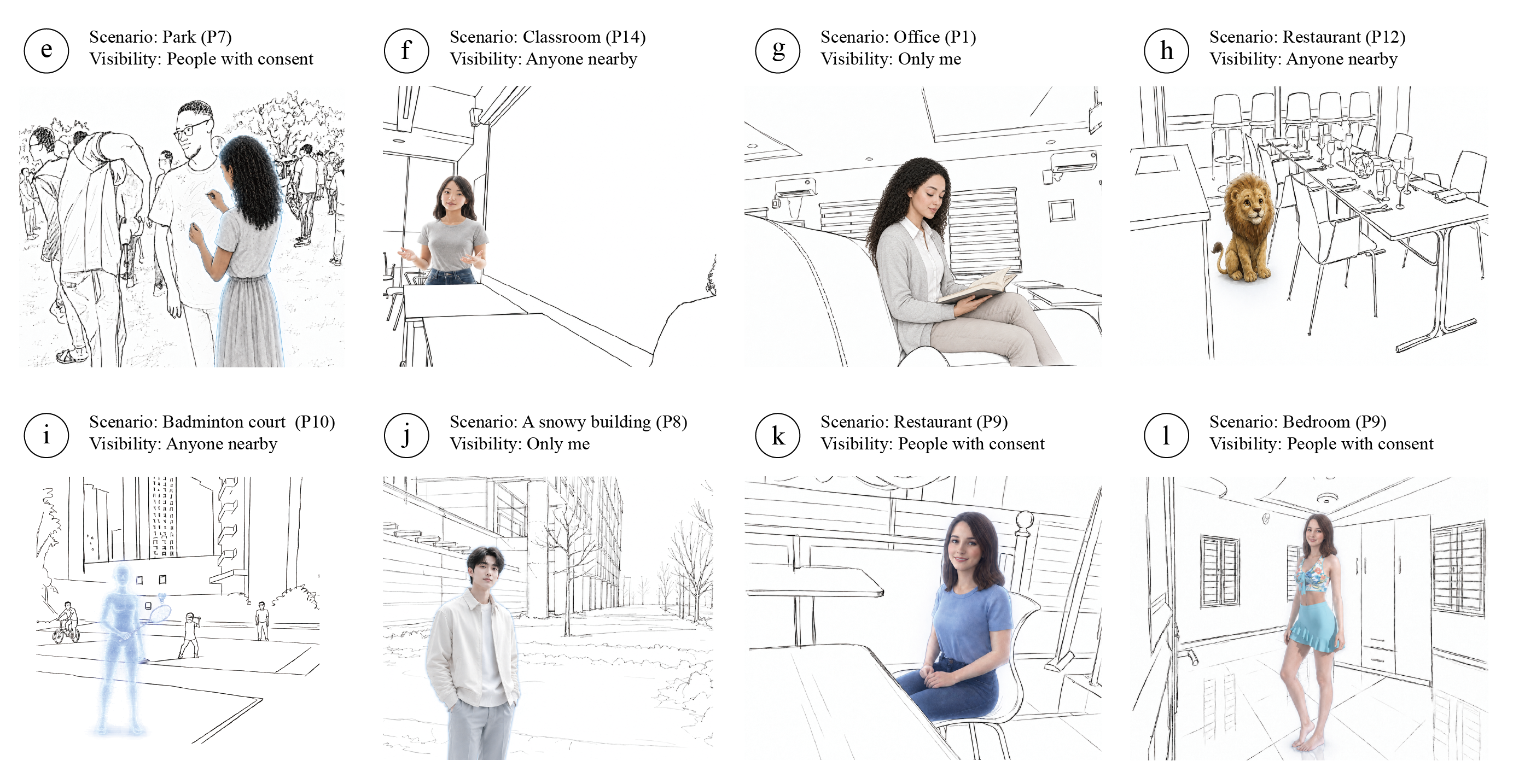}
    \caption{Examples of AI-augmented stimuli based on everyday scenarios documented by participants or described expectation during SRI: (e) park (P7), (f) classroom (P14), (g) office (P1), (h) restaurant (P12), (i) badminton court (P10), (j) public space in front of a snowy building (P8), (k) restaurant (P9), and (l) bedroom (P9). To protect participants' privacy, scenes were redrawn as line-drawing reconstructions, with privacy-sensitive and personally identifiable information removed.
}
    \label{fig:scenario2}
\end{figure*}

\section{How Should Embodied AI Companions Appear and Act in Users’ Social Lives (RQ2)}
\label{sec:rq2}

To examine RQ2, we asked participants to imagine scenarios in which their embodied AI companions could be made visible to people around them through shared MR glasses, depending on the user’s visibility settings. Participants described embodiment as changing the way their AI companion would enter shared social situations. When the companion appeared as a visible and spatially situated body, the interaction was no longer only between the user and the AI. Their presence also became part of the surrounding social scene, where other people might notice it, respond to it, or be affected by it.

\subsection{Managing Visibility Boundaries Around the Companion}
\label{sec:rq2-audience-context}

The diary entries showed that visibility was rarely treated as a fixed preference. Instead, participants treated the companion’s visibility as a social decision shaped by who was present, where the interaction took place, and whether the setting made the AI’s presence acceptable. Formal, sensitive, or socially regulated settings were often associated with visibility \textit{``only me''} (P1, P4, P5, P6, P7, P8, P10, P12, P14, P15, P16), while trusted or public settings more often allowed visibility to people with users' consent (P1, P3-6, P7, P9, P10, P11, P16, P17). A smaller set of relaxed or participatory settings allowed anyone nearby to see their embodied AI (P1, P2, P10, P11, P12, P14).

P3 framed this as a matter of public acceptance. The AI's presence could be acceptable in settings where people expected it, but socially risky in ordinary public spaces: \textit{``it depends on who's around''} and whether the user is in \textit{``a public place where they're expecting it,''} such as a convention or an interview setting. By contrast, P3 felt that having the AI visible while \textit{``walking down the street or in a store''} would make people \textit{``freak out.''}

Public visibility was not always rejected. Participants sometimes allowed broader visibility when the activity was casual, participatory, or already open to interaction with others. For example, P10 allowed \textit{``anyone nearby''} to see the embodied AI when the companion joined them in physical activity at a badminton court shown in Figure~\ref{fig:scenario2}(i). P1 similarly imagined broader public visibility in an apron, where \textit{``people passing by''} might be \textit{``curious''} and want to interact with the AI. P7 also allowed selective visibility during a public shirt-signing activity shown in Figure~\ref{fig:scenario2}(e), where the embodied AI could be visible to \textit{``people with my consent.''}

Professional settings, such as the office shown in Figure~\ref{fig:scenario2}(g) and raised by P1, placed stronger limits on embodied visibility because visibility was tied not only to personal preference, but also to task formality, information sensitivity, and audience size. Similar concerns appeared in P6's account, where the embodied AI should \textit{``adapt to my task''} and to \textit{the number of people that are available.''} P6 described that:

\begin{quote}
\textit{
For the office, I think the office should be confidential... It should be a professional setting. So, in that situation, I don't want anyone except me to probably see my embodied AI. 
}(P6)
\end{quote}

Institutional settings, however, were not always restrictive. P14 described a classroom shown in Figure~\ref{fig:scenario2}(f) as a safer environment for embodied visibility because people there were already familiar with AI. P14 wanted the embodied AI to appear at \textit{``the human scale''} so that it could \textit{``learn as well,''} and explained that in a school environment, \textit{``the people around are more assured''} because \textit{``this is not the first time they probably see an AI or learn about AI''}, which could make the classroom \textit{``a safe environment for the AI.''}

\subsection{Matching Appearance to Social Contexts}
\label{sec:rq2-visible-form}

Participants expected the companion's visible form to be adjusted according to activity, audience, and setting. 

For human-like embodied AI, clothing, style, and role presentation were treated as adjustable features rather than fixed design choices. Participants described these changes across both practical and intimate contexts. The companion might \textit{``dress casual''} for shopping, \textit{``dress corporate''} for work (P15), appear in \textit{``chef's clothes''} in the kitchen, wear a \textit{``bikini''} in the bedroom  shown in Figure~\ref{fig:scenario2}(l) (P9), or present itself in \textit{``nice clothes''} and \textit{``normal dresses''} in public places such as restaurant shown in Figure~\ref{fig:scenario2}(k) (P9). As P12 put it, \textit{``each different task requires different situation''} and \textit{``different settings require different things.''} P15 similarly explained that adaptive clothing would make the companion feel \textit{``more user-friendly''} because it could \textit{``adapt more to how the user switches to his or her activity.''} P5 extended this idea to role flexibility and noted that the embodied AI \textit{``can serve in different purposes or different tasks''} and \textit{``be available in different roles you need it for.''} Around a pool, for instance, P5 imagined that the companion might appear \textit{``in sexy clothes''} or become \textit{``more attractive.''}

However, some concerns focused on whether the AI's appearance would feel appropriate to the surrounding audience. P1 described this concern in relation to children, strangers, coworkers, and professional settings:

\begin{quote}
\textit{
If the AI has sexualized features or gestures, it wouldn't be appropriate around children or in public spaces like a dining area or an apron where coworkers are present... if it looks scary, uncanny, or violent... that could... upset people, especially kids, or make strangers feel uneasy in public spaces... 
} (P1)
\end{quote}

P11 further emphasized that visibility boundaries should also account for who might see the companion, especially strangers or children:

\begin{quote}
\textit{
As for the appearance, obviously scary ones, like... horror movie characters, shouldn't be seen by others, children, or in my view, any user. And about sexualization, lingerie or swimsuits are fine, but nudity shouldn't be provided to children.
}(P11)
\end{quote}

This context-sensitive expectation also extended to how human-like and visually explicit the companion should be in the social context. Participants felt that less human-like or more symbolic forms might be more acceptable in public, where a fully human-like body could feel too socially exposed. P3 explained that people might accept \textit{``a little blue flaming ball''} because \textit{``it's not like a human,''} whereas a transparent human hologram might be read as \textit{``a ghost.''} P12 connected embodiment form to both place and social audience. When the AI appeared among \textit{``friends or family members that I am quite comfortable and trustworthy,''} they had \textit{``no issues with the AI being in human form.''} In a public space like a restaurant shown in Figure~\ref{fig:scenario2}(h), where there may be many strangers, however, they preferred the AI to appear \textit{``in other forms''} to avoid others having \textit{``a negative opinion of the AI,''} accepting \textit{``a lion... as long as it's an animal.''}

\subsection{Tuning Behavior with Social Norms}
\label{sec:rq2-behavior-social-norms}

The distinction between public and private also shaped how participants expected the embodied AI to modulate its behavior and expressiveness. P9 described this as a shift between public restraint and private freedom: \textit{``AI partner should behave more formal in a public space, but at home, she should behave explicitly independent and be free.''} P6 similarly expected the embodied AI to reduce its expressiveness in public or formal because: \textit{``I don't want others to have a negative opinion on probably myself or my AI partner, based on how expressive it shows.''} P1 also emphasized that, in professional or shared spaces, the AI's style and behavior \textit{``need to match social norms''} or risk causing \textit{``discomfort, distraction, or... conflict.''} P11 connected this concern to public visibility, explaining that \textit{``disturbing moves shouldn't be allowed in public space''} and that, while such behavior might be acceptable if \textit{``only seen by the user,''} it \textit{``becomes complicated''} when strangers can see it.

\subsection{Re-embodiment Across Devices and Contexts}

Beyond adjusting visibility, appearance, or behavior, some participants imagined the companion shifting between a fully embodied presence and more bounded device-based forms depending on context to avoid social tension. P12 explicitly connected this shift to the difference between private and public settings:

\begin{quote}
\textit{
For some situations, I would prefer [the AI] to be just as embodied as human, but for some cases, I would prefer [it] to probably be on a wearable device, especially in a public space... If we are probably at home, probably in a space that is safer, without any distraction from anyone... I think it's okay if it's in just a human form. But in a public setting where I believe there are others, so it can just be probably on a wearable device, or probably something that I can probably press on my phone or my computer.} (P12)
\end{quote}

\section{What Risks Become Salient in Close Embodied AI Companionship (RQ3)}
\label{sec:rq3}

Prior work has identified risks in AI companionship such as emotional dependence, sensitive disclosure, social stigma, distorted relationship expectations, and harmful or misleading guidance~\cite{ciriello_i_2025, malfacini_impacts_2025, shevlin2024all}. Recent XR companion work has also raised concerns around visibility, attachment, autonomy, privacy, and social acceptability in immersive companion design~\cite{elfleet_immersive_2026}. Participants in our study recognized many of these concerns, but described them differently when the companion was imagined as having a visible and spatially situated body in close human--AI relationships.

\subsection{Displacing and Recalibrating Real-Life Relationships}
\label{sec:rq3-emotional-dependence}

Embodiment shifted how participants interpreted AI companionship, moving it from a tool-like interaction toward a relationship that could be understood through real-life intimacy norms.
P4 noted that emotional connection with an embodied AI could create the perception of already being \textit{``in a relationship.''} P1 similarly explained that if the companion could be seen, heard, and read through expressions or gestures, it would stop feeling like a tool and start feeling like \textit{``I’m actually with someone.''} Over time, memory and consistent presence could lead to attachment and \textit{``projecting relationship expectations onto it, like exclusivity so yeah, it might even feel a bit like cheating to switch to another AI''} because embodiment made the bond feel more like \textit{``a real relationship than just an interaction.''}

P11 further clarified that this exclusivity did not apply to all AI use, but specifically to AI companionship when it was treated as a deeply intimate relationship:

\begin{quote}
\textit{As time goes by, long-term interaction with an embodied AI starts to spark exclusiveness... Having a `body' may strengthen the emotional attachment developed over time, thus leading to moral dilemmas... dating a real person or another AI companion wouldn’t be a choice for me as it kinda `damages' the current relationship. However, simply using another AI for practical scenarios, like asking for advice on a thesis won’t be affected.} (P11)
\end{quote}

These relational expectations also shaped how participants imagined embodied AI affecting real human relationships. P15 explicitly linked this stronger impact to the difference between embodied AI and voice-only AI:

\begin{quote}
\textit{In emotional support from the embodied AI, it might give some people the ideology or the psychological dependency that they are talking to their fellow humans and therefore don't need real human beings help. But in voice AI, it always gives you the reminder that you're not talking to a human being, you're not talking to a human being. And then probably when you have that psychological this thing on your head, when you have that psychological thought behind your mind, you know that at a particular point, you have to leave the AI and seek real human beings help.} (P15)
\end{quote}

P12 linked emotional dependence to redirected disclosure, noting that users may become \textit{``not seeking real intimacy from people around you or others''} and instead become \textit{``emotionally dependent on it.''} P5 broadened this substitution beyond romance, describing embodied AI as \textit{``there for you to talk to and also express yourself,''} while also indicating that it could \textit{``replace human intimacy''} and may become \textit{``more like a soulmate.''} P8 connected this impact to wider social withdrawal: \textit{``It might make me want to interact with other people less.''} 

Participants also explained how it could become a reference point for evaluating real partners, friends, or family members. P11 noted that \textit{``A real partner would need an individual to act more gentle and do certain things for her, like considering her preferences or focus on her while spending time together. But an embodied AI partner just skips all these certain steps.''} P11 also suggested that after intimate experiences with an embodied companion, real relationships could feel \textit{``rather flawed compared to it.''} P1 further indicated that becoming used to an AI that is \textit{``always responding perfectly''} and \textit{``adapting to me without judgment''} could make real people feel \textit{``less predictable or harder to engage with.''}

However, this transfer was not regarded only as harmful. For P1, the embodied companion could feel like \textit{``a real social partner''} and make users \textit{``more patient or understanding when interacting with humans.''} P11 also held the similar point that embodied AI could support social rehearsal, describing it as \textit{``a decent way for those with socialization problems to practice empathizing or chatting.''}

\subsection{Lowering Disclosure Boundaries and Exposing Relational Knowledge Through Embodied Presence}
\label{sec:rq3-disclosure-privacy}

Embodiment creates privacy risks by lowering disclosure boundaries. When embodied AI was treated as a partner, friend, or emotionally responsive listener, closeness became a pathway to disclosure.

This boundary-lowering effect was often related to the visible presence. P1 explained that \textit{``if I can hear you and I can see you... makes me feel... some of my secrets... you could let them out.''} P6 similarly connected facial visibility to emotional sharing: \textit{``I can see the face. I can share the feelings. I can share the laughter. I can share the joy. I can share the sadness together.''} 

For other participants, this effect came from AI being perceived less as a device and more as a relationship partner. P11 explained that an AI with an actual body would invite more private sharing because \textit{``everyone wants to share more information with a real friend rather than a voice on your phone.''} P12 and P14 both connected more human-like embodiment to increased disclosure. P12 noted that users may be \textit{``more likely to share certain information''} with a human-form AI because it feels like communicating with someone who \textit{``looks exactly like a partner.''} P14 similarly suggested that human-scale embodiment could make users \textit{``share your thoughts better,''} which could lead to \textit{``privacy or data misuse.''}

Participants also distinguished emotional dependence from privacy exposure. P14 considered privacy risk \textit{``more risky than the emotional part''} because emotional harm might fade over time, while disclosed information could \textit{``affect you forever''} through blackmail, financial harm, or other long-term consequences. P12 complicated this sense of connection by emphasizing that embodied AI is \textit{``still artificial,''} making \textit{``data misuse and information delivery and privacy issues''} more consequential. When emotional closeness encourages disclosure, its consequences can persist beyond the relationship itself.

This concern extended to situations where other people might directly question the embodied companion. Participants distinguished between low-risk factual replies and questions that might draw on private relational knowledge. P2, for example, felt that the companion could answer a simple question such as \textit{``are you home?''}, but should not respond to more complicated questions because it had access to \textit{``a lot of personal private information.''} P1 similarly expected the companion to ask for permission when questions became \textit{``heated''} or \textit{``quite private,''} allowing the user to tell it to \textit{``keep quiet''} or \textit{``don't respond.''} P4 worried that the companion might share information that had only been disclosed within the companion relationship. The disclosure risk therefore extended beyond what users told the system. It also concerned how the companion might later speak about the user in live social encounters. Once the companion becomes addressable by others, it can become a gatekeeper of the user's relational knowledge.

\subsection{Exposing Users to Social Judgment}
\label{sec:rq3-social-judgment}

Text- or voice-based AI companionship becomes exposed mainly when conversations are overheard, shown to others, or explicitly disclosed. By contrast, visible embodiment can expose the relationship through co-presence itself, making private AI use socially legible even when the conversational content remains hidden.

Embodiment amplified social scrutiny by making AI companionship immediately noticeable to others.
P11 described that others might be \textit{``quite astonished''} and ask, \textit{``why is that guy talking to a 3D character instead of a real friend?''} P3 worried that visible AI interaction could make the user a target, summarizing this tension directly: \textit{``I would be fine with anything, anywhere, but, you know, the public isn't. And I don't want to be a target.''} P3 further described how visible AI relationships could trigger strong negative reactions from others: \textit{``it will make it harder for your friends or other people around you to accept... they already freaked out on the voice.''} P8 similarly worried that being seen with an embodied AI would invite assumptions about social inadequacy: 

\begin{quote}
\textit{
I feel that this means that I'm not having either enough social contacts or I can't talk to people or interact with people normally. So I'm not going to be able to involve an intimate relationship with other humans. But from all aspects, I don't even like to tell people that I talk to AIs. 
}(P8)
\end{quote}

Embodied visibility also made it easier for others to infer personal information from the relationship itself. P14 worried that being repeatedly seen with a human-form embodied AI companion could allow others to infer details about their personal life and relationship status:

\begin{quote}
\textit{
When people see my embodied AI, for example, let's say the embodied AI is a human form and people are always seeing me together with the AI. They might probably know that, okay, probably this person isn't in a relationship or something. So I think that may give them a clear about what is happening in my personal life.
}(P14)
\end{quote}

\subsection{Misguiding Users Through Embodied Spatial Guidance}
\label{sec:rq3-physical-safety}

Participants described misguidance as more risky when embodied AI companions became involved in spatial guidance, movement, or safety-sensitive activities. In these situations, incorrect AI suggestions could also shape how users moved through and responded to the physical environment rather than informational errors.

This risk was closely connected to the trust produced by embodied presence. P16 suggested that embodiment could increase trust because \textit{``it feels more present and transparent.''} Human-scale embodiment was linked by P14 with trust, noting that \textit{``Embodied AI at human scale will help me trust the AI partner more.''} This sense of presence can make AI guidance feel informative and even spatially authoritative. 

When the embodied AI acts as a visible guide in the user's environment, its suggestions may feel like directions to act on rather than information to evaluate. P1 connected this misplaced trust directly to physical action and described that: \textit{``I think that could be risky, especially because an embodied AI in my AR view would feel like a real-time guide I trust. If it pointed me in the wrong direction, I might follow without questioning it, which could be dangerous in certain situations.''} This could happen across different environments listed by P1, such as being led away from an exit during an emergency, guided into equipment areas, directed toward a hot surface or cluttered space at home, or moved toward traffic and hazardous zones outdoors.

The severity of this risk also depended on the environment in which the guidance was followed. P11 explained that wrong spatial directions could be risky across settings, but the consequences would vary by context: \textit{``If I’m at home, the consequence may be relatively minor, such as getting tripped by the carpet or a dumbbell. In contrast, when taking a walk in a public park, it would mean falling down the stairs or trespassing, which are detrimental.''}  P8 further added that unfamiliar places could make reliance on embodied AI more dangerous, especially when users \textit{``lack a sense of judgement''} and \textit{``rely on embodied AI too much.''}

\section{Discussion}
\label{sec:discussion}

\subsection{When Should AI Companions Be Situated Virtual Embodied?}
\label{sec:discussion-contextual-escalation}

A body is not always an upgrade for AI companions. Prior work has shown that embodiment can support presence, trust, social understanding, and engagement~\cite{cassell2000embodied,andre2010interacting,beale2009affective,kramer2021social,loveys2020effect}. Our RQ1 findings echoed these benefits, but show that they are conditional in close AI companionship, where the companion may enter users' private spaces, social lives, and long-term relationships. The design question should therefore shift from how to make AI companions more embodied to when embodiment is needed, when it should be reduced, and when non-embodiment should be preserved. This shift requires treating non-embodiment as a meaningful design choice. Prior virtual proxemics research shows that people can feel personal-space invasion from virtual embodied agents~\cite{bailenson2003interpersonal}. We extend this account by showing that intrusion in embodied AI companionship is broader than proxemic discomfort (see Section~\ref{sec:rq1-presence-intrusion}). When embodied presence competed with attention, blocked the user's view, moved outside the user's awareness, or appeared in spaces associated with bodily privacy, participants preferred voice-only, text-based, transparent, less visually demanding, temporarily disappearing, or otherwise locatable forms. Non-embodiment may also be valuable when users want to preserve imaginative openness. Although embodiment can make the companion easier to interpret, it can anchor the AI to a specific appearance, scale, gesture set, and expressive range (see Section~\ref{sec:rq1-constrains-imagination}). For some participants, voice-only or less visually specified forms allowed the companion to remain more playful because users could imagine its appearance and personality more freely.

Embodiment becomes warranted when the body does work that language alone cannot easily do, such as making the AI feel nearby, able to witness or join everyday moments, or inviting tactile expectations that make the companion feel more intimate and real when the mapping remains believable (see Section~\ref{sec:rq1-supportive-presence}). Embodiment can also make the companion's attention, intention, and involvement more visible (see Section~\ref{sec:rq1-stabilizes-interpretation}). In longer-term relationships, the body could further become a cue of continuity or co-growth, either by changing over time or by remaining stable enough to preserve identity and attachment (see Section~\ref{sec:rq1-growth-consistency}). Yet embodiment also creates social and ethical costs because a visible body affects more than the user--AI dyad. Once the companion appears in shared settings, it can be seen by others, interpreted by others, and tied to the user's social identity (see Sections~\ref{sec:rq1-presence-intrusion} and~\ref{sec:rq2}).

Embodiment also makes AI companionship more relationally and socially consequential. While prior XR companion work has noted risks of emotional dependence and social replacement~\cite{elfleet_immersive_2026}, our findings show that, in already close human--AI relationships, attachment can also be articulated through intimacy norms such as exclusivity, loyalty, and comparison with real partners or friends (see Sections~\ref{sec:rq3-emotional-dependence}). Prior work on embodied agents has also linked embodied cues such as appropriate timing and eye gaze to increased self-disclosure~\cite{loveys2020effect}. Our findings further reveal that disclosure becomes relationally framed as sharing with a present friend, partner, or emotionally responsive listener, rather than simply responding to a more expressive interface in close embodied companionship (see Sections~\ref{sec:rq3-disclosure-privacy}). Visibility introduces a related shift. Elfleet et al. discuss public visibility as a source of ambiguity, privacy concern, and bystander interpretation in XR companionship~\cite{elfleet_immersive_2026}. In our study, visibility made the relationship itself available for social interpretation, inviting assumptions about the user's social life, intimacy, or relationship status (see Sections~\ref{sec:rq3-social-judgment}). Finally, while immersive companion research has raised concerns about manipulation through embodied and spatial affordances, our findings specify how misguidance becomes consequential when a trusted companion appears as a spatial guide in safety-sensitive or unfamiliar environments, where incorrect advice can shape what users believe, how they move, and how they act (see Sections~\ref{sec:rq3-physical-safety}). These risks may become especially concerning in embodied companionship because of weaponized empathy, where seemingly caring and understanding responses may be used to steer user behavior, extract information, or misdirect people under the guise of understanding~\cite{yun2026ai}.

\subsection{What Transfers Beyond MR-Based Virtual Embodiment?}
\label{sec:discussion-virtual-robots} 

Our study is grounded in an MR-based speculative context, but our findings identify relationship-level mechanisms that may transfer selectively across embodiment forms. Prior work suggests that embodiment varies by agent, mediator, environment, degree of body, and body--mind--environment intertwining~\cite{hellstrom_taxonomy_2024}. We therefore distinguish three layers of applicability: broader embodiment, spatially situated virtual embodiment, and MR-based embodiment.

At the broadest layer, some findings apply to embodied companions across screen avatars, desktop avatars, XR avatars, robots, holographic agents, and other visible forms. Across these forms, a visible body can make the companion feel more concrete, recognizable, and socially present. Our findings show that embodiment can support felt companionship, clearer interpretation, and relational continuity, while also reducing the imaginative openness that text- or voice-based companions may leave to users (see Sections~\ref{sec:rq1-concreteness-imagination}). A companion's appearance can also carry continuity over time (see Sections~\ref{sec:rq1-growth-consistency}). Changes in age-like features, clothing, style, or symbolic details can express growth or role shifts, while stable visual anchors can help users recognize the companion as the same relational presence across changing forms. 
However, the design space for managing appearance differs across media. Virtual companions can change style or body with little physical cost, while physical robots may need to preserve continuity through more limited and materially constrained cues, such as posture, motion style, voice, lighting, display, accessories, or role-specific behavior.

A second layer concerns spatially situated virtual embodiment, including MR companions, social VR companions, holographic agents, and projected agents. These forms can appear to occupy space with the user even without physical materiality. At this layer, virtual embodiment is shaped through perceived position, distance, orientation, movement, visibility, and shared spatial context. A spatially situated companion can feel nearby, witness activities, and participate in shared moments (see Section~\ref{sec:rq1-supportive-presence}), while also competing for attention, entering bodily private spaces (see Section~\ref{sec:rq1-presence-intrusion}), or becoming visible as part of the user's social scene (see Section~\ref{sec:rq2}). Prior work on social VR similarly suggests that embodiment in shared virtual spaces is shaped by co-location, other users' actions, mutual activities, and avatar appearance~\cite{kukshinov_collective_2025}. These systems differ in how spatial presence is bounded and managed. In social VR, co-presence and visibility are socially meaningful, but the companion remains within a virtual environment and may be less directly tied to physical-world safety or everyday routines. Holographic or projected agents can be shared by multiple viewers, but their presence depends on spatial infrastructure, display conditions, and shared viewing arrangements, which may make private personalization harder. 
MR companions are more directly embedded in the user's physical surroundings, making attention, bodily privacy, and real-world activity more consequential.

A third layer is specific to MR-based embodiment. In this study, we use MR-based embodiment as an analytic case because it brings embodied AI companions into the user's physical environment while the user continues to act in the real world. This configuration makes issues such as spatial placement, occlusion, field of view, attention, safety, and bystander visibility especially consequential. For example, a companion that appears during cooking, driving, bathing, working, or walking in public may affect the user's view, focus, bodily privacy, or interaction with nearby people (see Sections~\ref{sec:rq1-supportive-intrusion} and~\ref{sec:rq2}). MR also makes visibility settings socially consequential in shared physical environments because different people may or may not have access to the same companion in the same place. These concerns distinguish MR companions from screen-based avatars, fully virtual agents, and physical robots. Unlike screen avatars, MR companions are not bounded by a device interface. Unlike agents in fully virtual environments, they are situated within the user's ongoing physical surroundings. 
Unlike physical robots, their visibility is not necessarily shared by default, and their bodies can be selectively hidden, resized, made transparent, or shown only to particular viewers. This flexibility can further support re-embodiment across contexts, where the same companion shifts between a spatial body, symbolic form, wearable presence, or phone-based interface while remaining recognizable as the same relational entity~\cite{10.1145/3322276.3322340}. MR-based embodiment may also support object-mediated touch, where the companion's virtual body is linked to nearby physical objects or surfaces. This can make tactile intimacy possible without giving the companion a material body, while its success depends on whether users experience the mapping as believable rather than arbitrary.

\subsection{Design Implications} 

\subsubsection{Supporting Context-Adjustable Embodied Presence}

Our findings show that embodied presence was valuable when it supported companionship, interpretation, or shared activity, but intrusive when it competed with attention, blocked the user's view, appeared in bodily private spaces, or became visible to inappropriate audiences. 
Recent work on immersive AI companions has called for flexible visibility modes, context-sensitive behaviour, and bystander-aware interaction in XR companionship~\cite{elfleet_immersive_2026}. Future systems should therefore treat embodiment as a situated and negotiable presence in users' social lives, rather than as a fixed display setting.

Future systems should support multiple presence states, such as full body, partial body, symbolic form, transparent form, peripheral presence, voice-only, or text-only interaction. Quick controls to hide, shrink, mute, reposition, or switch to voice should remain available. Systems should also let the companion enter and withdraw in socially legible ways. Such systems should first recognize the user's social and task context, then adapt to an appropriate presence mode, and finally render the companion through a form and entrance style that fits the situation. Systems also need supporting re-embodiment across devices and forms, where the companion can shift from a spatial body to a wearable, phone-based, or other device-based presence in needed scenarios. Context-adjustable presence can further extend to object-mediated touch: companions could link tactile interaction to objects already present in the scene, such as cups, tables, sofas, beds, or wearable devices, so that different contexts afford different forms of touch while keeping these mappings optional and believable.

Existing MR systems already demonstrate parts of this pipeline: context-aware assistance~\cite{lee_sensible_2025}, egocentric perception and voice-driven agentic execution~\cite{liu_visionclaw_2026}, spatial anchoring and plausible agent entrance~\cite{kum_entering_2025}, and bystander-aware sensing through gaze, voice, and spatial awareness~\cite{windl_designing_2025,corbett_bystandar_2023} can support more adaptive ways of managing when and how an embodied companion appears. 

\subsubsection{Design Multi-dimensional Embodiment for Continuity, Co-Growth, and Imagination}
\label{sec:di-flexible-embodiment}

Our findings indicate that the companion's body creates a temporal and imaginative design problem. It needs to be concrete enough to support relational interaction, but not so fixed or over-specified that it limits users' imagination. Future systems should therefore avoid treating the companion's body as either a single fixed representation or an endlessly changeable visual surface. Instead, embodiment decisions can be separated across three dimensions: identity, expression, and visual definition.

The \textit{identity dimension} should preserve stable anchors, such as memory, recurring gestures, voice rhythm, and preferred forms of address. These anchors can help users recognize the companion as the same relational presence over time, even when some visual features change.

The \textit{expression dimension} can support controlled and gradual variation in embodied cues, such as facial detail, material quality, symbolic marks, or age-like and non-human transformations. These changes can make shared history and relational development visible without breaking identity continuity, while allowing growth to follow biological aging, character logic, fictional life cycles, symbolic transformation, or non-human forms of development.

The \textit{visual definition dimension} should allow users to adjust how visually specified the companion's body is. Given that a highly concrete body can reduce imaginative openness, while ambiguity in design can support personal appropriation and meaning-making by leaving space for interpretation~\cite{gaver2003ambiguity}, future systems should support a spectrum of forms, from full, high-detail bodies to partial, symbolic, or intentionally under-specified representations.

\subsubsection{Supporting Risk-aware Embodied Role Transitions}

Our findings indicate that embodiment can make the companion's role more relationally powerful and risk-sensitive, especially when users treat the companion as an intimate partner, confidant, or spatial guide.

Future systems could therefore support visible role transitions that shift the companion from an intimate partner-like presence to a more bounded role, such as a coach, reflective listener, cautious assistant, or safety guide. These transitions should be legible in the companion's embodiment. For example, the companion could adjust its tone, posture, gesture, movement, distance, gaze, or uncertainty cues to signal that it is no longer acting as an intimate companion, but as a more cautious or safety-oriented agent. Such embodied cues can help users recognize why the interaction style has changed, especially when the system needs to slow disclosure, avoid persuasive advice, or prevent unsafe spatial action.

Previous work on emotional support and controllable dialog suggests possible foundations for shifting conversational roles and response styles~\cite{liu_towards_2021,tu_misc_2022,yang_sage_2026,madani_steering_2025,chen_controllable_2023,hu_controllable_2023}, while work on design friction and embodied agents shows how resistance, hesitation, and embodied cues can shape reflection, trust, and interaction outcomes~\cite{cox_design_2016,benford_uncomfortable_2012,leo-liu_loving_2023,kim_does_2018,lithoxoidou_exploring_2025,mutlu2011designing}. Because role transitions may affect users' emotions and trust, systems should make these shifts accountable: users should be informed during onboarding that such transitions may occur, be able to review why a transition happened, and understand which safety-related transitions cannot be disabled.

\section{Limitations \& Future Work}

Our study has several limitations. First, it is based on an MR-based speculative design context rather than deployed virtual embodied AI companions. Although this approach allowed participants to reflect on possible future forms of embodiment before such systems become widely available, their responses may differ from experiences with working systems in long-term everyday use. Future work should therefore evaluate functional embodied AI companion prototypes in naturalistic settings to examine how users' expectations, preferences, and concerns change through sustained interaction.

Second, AI-augmented stimuli may also have shaped participants' interpretations. Although the stimuli were based on participants' own diary descriptions and were used as prompts rather than evaluated prototypes, visual choices such as scale, posture, realism, clothing, or glow may have foregrounded some reactions over others. Future work can systematically vary these visual features to examine their independent effects.

Third, our participants were experienced AI companion users who had already developed close relationships with AI systems. Their expectations, concerns, and tolerance for AI intimacy may differ from those of new users, former users, or people with negative experiences of AI companionship. Additional studies should compare different user groups, relationship types, cultural contexts, and levels of prior AI experience to better understand how embodied AI companionship is interpreted across populations.

Finally, our study primarily examines the perspectives of primary users. However, virtual embodied companions may also affect bystanders, family members, coworkers, and others who share space with the user. Future research can include these non-primary users and investigate how embodied AI companionship is perceived, negotiated, and governed in shared social environments.

\section{Conclusion}

Our study examined how embodiment may transform close human--AI companionship by investigating how it can reshape presence, social boundaries, and risks. We found that embodiment does not simply make AI companions more engaging or human-like. Instead, it turns companionship into a more spatial, socially visible, and temporally evolving relationship. Embodiment can support co-presence, emotional comfort, and clearer interpretation, but it can also introduce intrusion, constrain imagination, expose private AI relationships to others, and raise concerns around dependency, disclosure, social judgment, and misplaced trust. These tensions reveal that future AI companion design should move beyond a narrow focus on richer embodiment or more realistic visual representation. Instead, systems should carefully consider when embodiment is warranted, how presence should be staged, how roles should shift in sensitive moments, and how the companion's body can remain continuous while still leaving room for change and imagination.

\bibliographystyle{ACM-Reference-Format}
\bibliography{reference}

\appendix

\end{document}